\documentclass[prl,twocolumn,showpacs,amsmath,amssymb]{revtex4}

\usepackage{graphicx}
\usepackage{dcolumn}
\usepackage{bm}
\usepackage{epsfig}

\begin{document}

\newcommand{\ep}{\varepsilon}
\newcommand{\up}{\uparrow}
\newcommand{\dn}{\downarrow}
\newcommand{\vectg}[1]{\mbox{\boldmath ${#1}$}}
\newcommand{\vect}[1]{{\bf #1}}

\title{Spin Hall Current Driven by Quantum Interferences in Mesoscopic Rashba Rings}

\author{Satofumi Souma and Branislav K. Nikoli\' c}
\affiliation{Department of Physics and Astronomy, University
of Delaware, Newark, DE 19716-2570}

\begin{abstract}
We propose an all-electrical nanoscopic structure where {\em pure} spin current is 
induced in the transverse voltage probes attached to {\em quantum-coherent} one-dimensional 
ring when conventional unpolarized charge current is injected through its longitudinal 
leads. Tuning of the Rashba spin-orbit coupling in semiconductor heterostructure 
hosting the ring generates quasi-periodic oscillations of the predicted spin Hall 
current due to {\em spin-sensitive quantum-interference effects} caused by the difference in 
Aharonov-Casher phase acquired by opposite spins states traveling clockwise and counterclockwise. Its 
amplitude  is comparable  to the mesoscopic spin Hall current predicted for finite-size two-dimensional 
electron gases,  while  it gets reduced in wide two-dimensional or disordered rings.
\end{abstract}

\pacs{72.25.Dc, 03.65.Vf, 03.65.Yz, 73.23.-b}
\maketitle

{\it Introduction}.---The increasing interest in spin-based
information processing has fomented the field of {\em semiconductor 
spintronics}~\cite{spin_book} where a plethora of concepts,  exploiting 
fundamental quantum phenomena that involve   electron spin, have 
arisen in order to generate and measure {\em pure spin currents}.  In contrast to 
conventional charge currents or spin-polarized charge currents, which have 
been  explored and utilized in  metallic spintronics over the past two 
decades~\cite{maekawa}, pure spin currents emerge when equal number of 
spin-$\uparrow$ and  spin-$\downarrow$  electron move in the opposite direction  so 
that net charge current is zero~\cite{guo_spin_current}. Early~\cite{extrinsic_1}  and recent~\cite{extrinsic_2,bauer,pareek,bhat,mucciolo} theoretical analysis has 
found potential sources of such current in: metallic or semiconductor paramagnets 
with spin-orbit (SO) dependent scattering on impurities (supporting extrinsic 
spin  Hall effect~\cite{extrinsic_1,extrinsic_2} as transverse spin 
current in response to longitudinal charge transport, or skew-scattering 
effects in $Y$-shaped semiconductor junctions~\cite{pareek}); multiprobe ferromagnet-normal metal hybrid devices~\cite{bauer}; optical injection in clean semiconductors~\cite{bhat}; and adiabatic 
spin pumping in mesoscopic systems~\cite{mucciolo}. Moreover, spin currents without accompanying charge current have been generated and detected in optical pump-probe experiments~\cite{stevens2003} and semiconductor quantum spin pumps~\cite{spin_battery}.

Recent theoretical hints at the existence of {\em intrinsic}
spin Hall effect in {\em clean} hole-doped~\cite{murakami}  or electron-doped~\cite{sinova}  
semiconductor  systems governed by SO couplings,  where pure transverse spin current (substantially 
larger than in the case of extrinsic effect) is  predicted as a response to longitudinal applied electric 
field,  has attracted considerable attention. This is essentially a semiclassical effect in 
which current $j_y^z$ of $z$-polarized spins flows along the $y$-axis
within an infinite clean homogeneous semiconductor system  penetrated by an
external macroscopic electric field $E_x$ along the $x$-axis. That is, it can
be explained using a wave packet formalism~\cite{sinova_bloch} where current
is generated by the anomalous velocity due to the Berry curvature of the Bloch
states in SO coupled systems, rather than the displacement of the electron
distribution function (as is the case of traditional charge currents 
accompanied by Joule heating). The generation and control of  pure  spin Hall current 
(that would be accompanied only by low-dissipative longitudinal charge current) could 
make possible spin manipulation without  magnetic fields or problematic 
coupling of ferromagnetic electrodes to semiconductors devices~\cite{rashba_review}.

The {\em non-equilibrium} spin current represents transport of spins between
two locations in real space. However, intense theoretical striving to understand 
the  nature  of intrinsic spin Hall current, quantified by $j_y^z$~\cite{spin_operator} 
and spin Hall conductivity  $\sigma_{sH} = j_y^z/E_x$,  suggest that $j_y^z \neq 0$  
might  not imply real transport of spins since in  dissipationless  transport regime through a 
clean  system it can be interpreted  as an  equilibrium background spin current existing even in the 
absence of  any external  electric field~\cite{rashba_eq}. In addition, studies concerned with 
the influence of disorder (spin-independent scattering off static impurities) on spin Hall effect~\cite{inoue}, 
as well as reexamination of the original arguments for clean systems~\cite{rashba_spin_hall}, converge 
toward the conclusion that $\sigma_{sH} \rightarrow 0$ in an infinite homogeneous two-dimensional electron 
gas (2DEG) with Rashba SO interaction~\cite{rashba} (such SO coupling is pertinent to 2DEG since it stems 
from the inversion asymmetry of the quantum well confining electric potential). Nevertheless, quantum transport analysis of {\em measurable}~\cite{stevens2003,meso_spin_hall} spin-resolved charge currents $I_p^\uparrow$, $I_p^\downarrow$ and corresponding spin currents $I_p^s = \frac{\hbar}{2e}(I_p^\uparrow - I_p^\downarrow)$ in 
the ideal leads (without SO interaction) of multiprobe Hall bars accessible to experiments predicts that 
a type of spin Hall current will appear in the transverse voltage probes~\cite{meso_spin_hall,sheng,ewelina} attached to a finite-size 2DEG with Rashba SO interaction. This is due to the fact that spin currents in
both the diffusive and the ballistic regime can be facilitated by macroscopic inhomogeneities~\cite{rashba_spin_hall}. Furthermore, possible signatures of spin Hall effect have 
been detected in finite-size 2D hole gases~\cite{wunderlich}.
\begin{figure}
\includegraphics[scale=0.35]{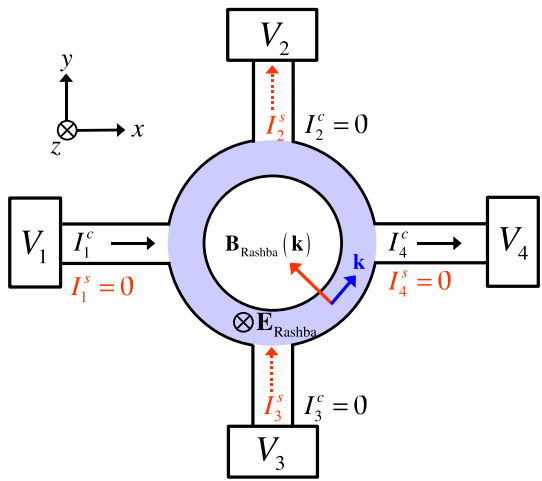}
\caption{The mesoscopic circuit serving as a generator of  the 
pure ($I_2=I_2^\uparrow+I_2^\downarrow=0$) spin Hall current $I_2^s=\frac{\hbar}{2e}(I_2^\uparrow-I_2^\downarrow) = -I_3^s$ in the transverse voltage probes ($V_2=V_3 \neq 0$, $I_2=I_3=0$) attached to
a ring realized using 2DEG in a semiconductor heterostructure~\cite{ring_2deg}. The injected 
unpolarized ($I_1^s=0$) current through (single-channel) longitudinal leads  
is subjected to the Rashba SO interaction  (nonzero in the shaded ring region), which acts as   
a momentum-dependent pseudomagnetic field ${\bf B}_{\rm Rashba}({\bf  k})$ arising due to the electric  field ${\bf E}_{\rm Rashba}$ confining the electrons to 2DEG.}\label{fig:ring_setup}
\end{figure}
\begin{figure}
\includegraphics[scale=0.7]{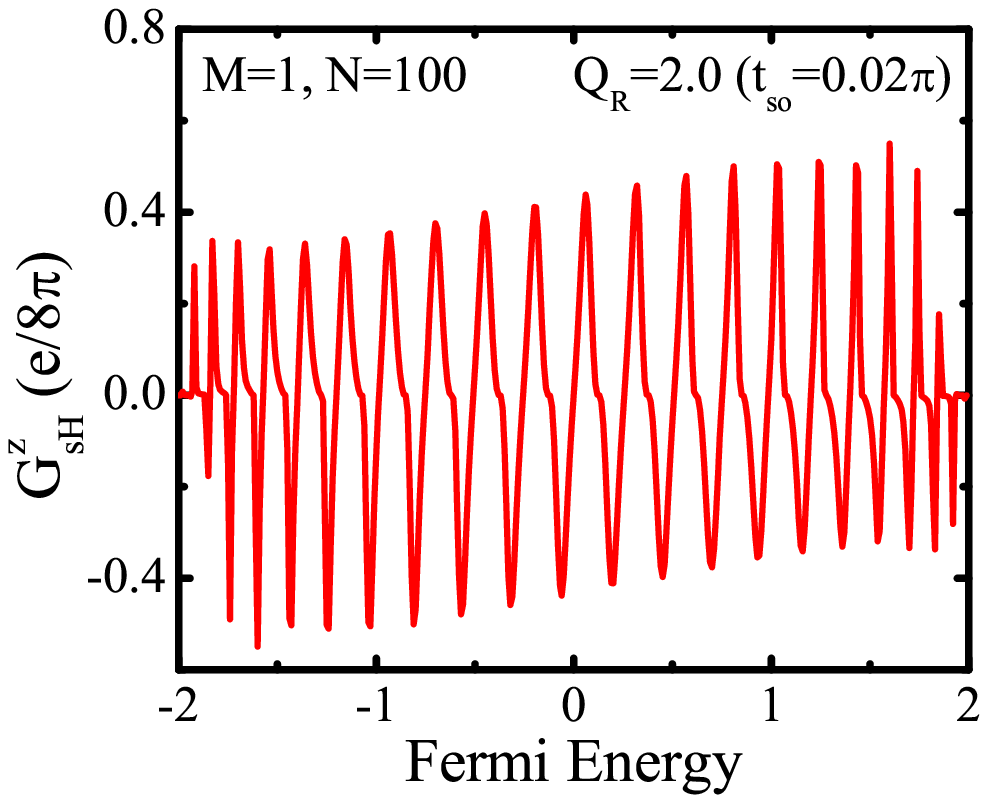}\vspace{-0.3in}
\includegraphics[scale=0.90]{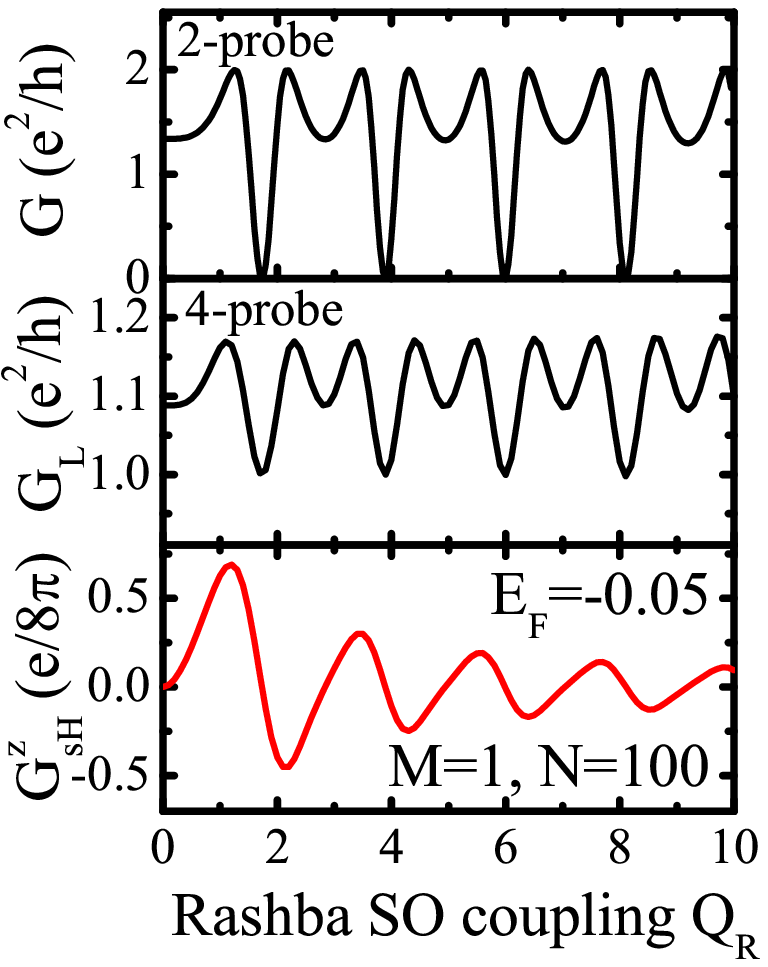}
\caption{The spin Hall conductance $G_{sH}^z$ (corresponding
to the detection of the $z$-component of pure spin current $I_2^s$)
for 1D ring ($M=1$, $N=100$ lattice sites around the ring) attached 
to four single-channel leads as a function  of the Fermi energy $E_F$ 
(upper panel) and the dimensionless Rashba SO coupling $Q_{\rm R} \equiv 
(\alpha/2at_0)N/\pi$ (lower panel). The lower panel  also plots the 
charge conductance $G(Q_{\rm R})$ of the corresponding two-terminal AC 
ring~\cite{diego,ac_ring}  as well as the longitudinal charge conductance 
$G_L (Q_{\rm R}) = I_4 /(V_1-V_4)$ of our four-terminal Rashba ring 
depicted in Fig.~\ref{fig:ring_setup}.}\label{fig:ring_1D}
\end{figure}

\begin{figure}
\includegraphics[scale=0.75]{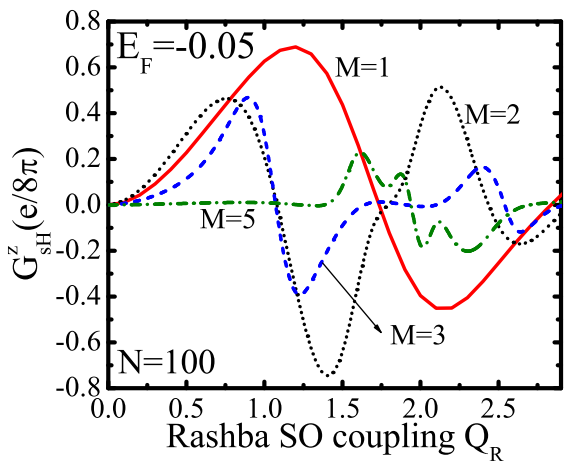}
\caption{The modulation of spin Hall conductance $G_{\rm sH}^z$ by 
changing the Rashba SO coupling  $Q_{\rm R} \equiv (\alpha/2at_0)N/\pi$ in 2D 
ballistic  rings of finite width (modeled by  $M \ge 1$ coupled concentric 1D ring chains
of $N=100$ lattice sites) attached to four single-channel leads. The
unpolarized current injected through the longitudinal leads is composed
of spin-$\uparrow$ and spin-$\downarrow$ electrons at the Fermi energy $E_{\rm F}=-0.05$.}\label{fig:ring_2D}
\end{figure}

Thus, it becomes intriguing to pose two fundamental questions: Is it possible to induce spin Hall current in {\em strictly one-dimensional} systems with {\em no bulk}? Does quantum coherence (i.e., {\em spin-interference effects}) play any role in mesoscopic spin Hall current induction that can leave unique experimentally observable signatures? In this letter we undertake answering both of these questions by analyzing the  spin-charge quantum  transport in the presence of Rashba SO coupling within mesoscopic ring-shaped conductor (realized using 2DEG in  semiconductor heterostructure~\cite{ring_2deg}), which is modeled by the following single-particle effective mass Hamiltonian 
\begin{eqnarray} \label{eq:rashba_hamil}
\hat{H} = \frac{\hat{\vect{p}}^2}{2m^*} +
\frac{\alpha}{\hbar}\left(\hat{\vectg{\sigma}}\times\hat{\vect{p}}\right)_z
+ V_{\rm conf}(x,y) + V_{\rm dis}(x,y).
\end{eqnarray}
Here $\hat{\vectg{\sigma}}$ is the vector of the Pauli spin
operator, $\hat{\vect{p}}$ is the momentum vector in  2D space, $\alpha$ is 
the strength of the Rashba SO coupling~\cite{rashba}, and  $V_{\rm conf}(x,y)$ is 
the potential  which confines electrons to a finite ring region. Such {\em Rashba ring}, attached to 
two  longitudinal current probes and two transverse voltage probes  (Fig.~\ref{fig:ring_setup}), 
will  generate spin Hall current in the transverse leads. As demonstrated in Fig.~\ref{fig:ring_1D} 
for 1D and in Fig.~\ref{fig:ring_2D} for 2D rings, which are free of disorder $V_{\rm dis}(x,y)=0$, the spin 
Hall  conductance  $G_{sH}^z = I_2^s/(V_1 - V_4)$ measuring the magnitude of  the transverse pure spin 
current in mesoscopic structures~\cite{meso_spin_hall,sheng,ewelina} will exhibit quasi-periodic oscillations, 
due to  spin quantum-interference effects, when Rashba SO coupling is increased (e.g., via gate electrode covering the ring~\cite{nitta}).

The ring conductors smaller than the dephasing length $L_\phi \lesssim 1\mu$m
(at low temperature $T \ll 1$K) have played an essential role in observing how
{\em coherent superpositions} of quantum states (i.e., quantum-interference effects) on mesoscopic scale leave imprint on measurable transport properties. That is, they represent a solid state realization of a two-slit experiment---an electron entering the ring can propagate in two possible directions (clockwise and counterclockwise) where superpositions of corresponding quantum states are sensitive to the acquired topological phases in magnetic [Aharonov-Bohm (AB) effect] or electric [Aharonov-Casher (AC) effect for particles with spin] external field whose 
changing generates an oscillatory pattern of the ring conductance~\cite{ring_2deg}. Moreover, recently 
proposed {\em all-electrical} mesoscopic spintronic 1D ring device~\cite{nitta_ring} would utilize the difference between AC phases of opposite spin states traveling clockwise and counterclockwise around the ring in a way in which their {\em spin interferences} will modulate the conductance of unpolarized charge current injected through single-channel leads between 0 and $2e^2/h$ by changing the Rashba electric field~\cite{diego,ac_ring}.

{\em Quantum transport of spin currents in 4-terminal Rashba rings}.---The charge 
currents  in mesoscopic structures attached to many leads are described by 
the multiprobe Landauer-B\" uttiker formulas~\cite{buttiker} 
\begin{equation}\label{eq:buttiker}
I_p = \sum_{q \neq p} G_{pq}(V_p - V_q),
\end{equation}
while the  analogous formulas for the spin currents in the leads are straightforwardly extracted from them~\cite{pareek,meso_spin_hall}
\begin{equation} \label{eq:spinbuttiker}
I_p^s =
\frac{\hbar}{2e}\sum_{q \neq p} (G_{qp}^{\rm out} V_p - G_{pq}^{\rm in} V_q).
\end{equation}
Here $G_{pq}^{\rm in}=G_{pq}^{\uparrow\uparrow}+G_{pq}^{\uparrow\downarrow}-G_{pq}^{\downarrow\uparrow}-G_{pq}^{\downarrow\downarrow}$ and $G_{qp}^{\rm out}=G_{qp}^{\uparrow\uparrow}+G_{qp}^{\downarrow\uparrow}-G_{qp}^{\uparrow\downarrow}-G_{qp}^{\downarrow\downarrow}$ have transparent physical interpretation: $\frac{\hbar}{2e}G_{qp}^{\rm out} V_p$ is the spin current flowing from the lead $p$ with voltage $V_p$ into other leads $q$ whose voltages are $V_q$, while $\frac{\hbar}{2e} G_{pq}^{\rm in} V_q$ is the spin current flowing from the leads
$q \neq p$ into the lead $p$ (the standard charge conductance coefficients are expressed in terms of the spin-resolved conductances as $G_{pq}=G_{pq}^{\uparrow\uparrow}+G_{pq}^{\uparrow\downarrow}+G_{pq}^{\downarrow\uparrow}+G_{pq}^{\downarrow\downarrow}$~\cite{purity}). The linear response conductance coefficients are related to the transmission  matrices ${\bf t}^{pq}$  between the leads $p$ and $q$ through the Landauer-type formula $G_{pq}^{\alpha \alpha^\prime}=\frac{e^2}{h} \sum_{i,j=1}^{M_{\rm leads}} |{\bf t}^{pq}_{ij,\alpha \alpha^\prime}|^2$, where 
 $|{\bf t}^{pq}_{ij,\alpha \alpha^\prime}|^2$ is the probability for spin-$\alpha^\prime$  electron incident in lead $q$ to be transmitted to lead $p$ as spin-$\alpha$ electron and $i,j$ label the transverse propagating modes (i.e., conducting channels) in the leads. The general expression for the spin Hall conductance is~\cite{meso_spin_hall}
\begin{equation} \label{eq:gh_explicit}
G_{sH}=\frac{\hbar}{2e} \left[(G_{12}^{\rm out}+G_{32}^{\rm out}+ G_{42}^{\rm out})\frac{V_2}{V_1} - G_{23}^{\rm in}\frac{V_3}{V_1} - G_{21}^{\rm in} \right],
\end{equation}
where we choose the reference potential to be $V_4=0$. We emphasize that, in general, there are three non-zero spin conductances corresponding to three components of the polarization of transported spin~\cite{meso_spin_hall}. For simplicity, we analyze only the $z$-component (i.e., we set the spin quantization axis for $\uparrow$, $\downarrow$ in Eq.~(\ref{eq:gh_explicit}) to be the $z$-axis).

We recall that Landauer transport paradigm spatially separates single-particle coherent and many-body inelastic processes by attaching the sample to huge electron reservoirs where, in order to simplify the scattering boundary conditions, semi-infinite ideal leads with vanishing spin and charge
interactions are inserted between the reservoirs and the scattering region.
Thus, even in the ballistic regime dissipation effects establishing steady
state transport are always incorporated, in contrast to the artifacts of the Kubo 
formalism which maps the intrinsic spin Hall current in an infinite dissipationless 
system driven by the electric field to an equivalent system containing only equilibrium spin 
currents~\cite{rashba_eq}. Here we clarify that apparent equilibrium solutions of the multiprobe 
spin current relations Eq.~(\ref{eq:spinbuttiker}), $V_q={\rm const.} \Rightarrow I_p^s \neq 0$ found 
in Ref.~\cite{pareek} to originate from   $G_{pq}^{\alpha \alpha^\prime} \neq G_{pq}^{\alpha -\alpha^\prime}$, actually {\em do not exist}. When all leads are at the same potential, a purely equilibrium non-zero term $\frac{\hbar}{2e} (G_{pp}^{\rm out} V_p - G_{pp}^{\rm in} V_p)$ (omitted in Ref.~\cite{pareek}) becomes 
relevant for $I_p^s$, canceling all other terms in Eq.~(\ref{eq:spinbuttiker}) to ensure that no 
{\em unphysical}  $I_p^s \neq 0$ would exist in the leads of an unbiased ($V_q$=const.)  mesoscopic structure.

The stationary states of a system 1D ring + two 1D leads can be found exactly  
by matching the wave functions in the leads to the eigenstates of the ring 
Hamiltonian Eq.~(\ref{eq:rashba_hamil}), thereby allowing one to obtain the 
charge conductance from the Landauer transmission formula~\cite{diego}. However, attaching two 
extra leads in the transverse direction, as well the finite width of the ring and/or  
presence of disorder within the ring region, requires to switch from wave function to 
some type of Green function formalism. Here we employ the real$\otimes$spin space Green 
function technique~\cite{meso_spin_hall,purity} which yields the {\em exact} (within single-particle 
picture) transmission  matrices ${\bf t}^{pq}$  between the leads $p$ and $q$. 
The computation of  the non-perturbative retarded Green function can be done efficiently using a local orbital basis representation 
of the Hamiltonian Eq.~(\ref{eq:rashba_hamil}), which we have introduced in Ref.~\cite{ac_ring} as a set of 
$M$ concentric chains composed of $N$ lattice sites spaced at a distance $a$. The characteristic energy scales 
of such lattice Hamiltonian are: the hopping between neighboring sites $t_0 = \hbar^2/(2 m^* a^2)$ (all energies will be measured in the units of $t_0$), and the Rashba hopping $t_{\rm so} = \alpha/2a$. It is also useful to measure the strength of the Rashba SO coupling within the ring region using a dimensionless parameter $Q_{\rm R} \equiv (t_{\rm so}/t_0)N/\pi$~\cite{diego,ac_ring}. Since the contact between the ring and the leads can be controlled precisely 
using a quantum point contact to ensure that unpolarized current is injected through a single open conducting channel, we assume 1D electrodes ($M_{\rm leads}=1$) while allowing for both strictly 1D rings $M=1$ and 2D rings of finite width $M > 1$ \cite{ac_ring}.

{\em Spin-interference effects in spin-Hall conductance}.---The rapid
oscillations of $G_{sH}^z(E_F)$ in Fig.~\ref{fig:ring_1D} arise due to
discrete nature of the energy spectrum in an isolated ring (note that once 
the leads are attached these eigenlevels acquire a finite width since electrons spend
finite time inside the ring before escaping into the leads). The charge conductance of the 
two-probe  1D AC ring~\cite{nitta_ring,diego,ac_ring} becomes zero at specific values 
of $Q_{\rm R}^{\rm min}$ for which  {\em destructive} spin-interference of opposite spins 
traveling in opposite directions around the ring takes place. For example, in a simplified 
treatment~\cite{diego} $G=\frac{e^2}{h}[1+\cos (\Phi_{\rm AC}^\up-\Phi_{\rm AC}^\dn)/2]$, where  $\Phi_{\rm AC}^\sigma=\pi(1+\sigma\sqrt{Q_{\rm R}^2+1})$ is the AC phase acquired by a spin-$\uparrow$ or spin-$\downarrow$ electron ($\sigma=\pm$ for $\up,\dn)$, has minima $G(Q_{\rm R}^{\rm min})=0$ at $Q_{\rm R}^{\rm min} \simeq\sqrt{n^2-1}$ ($n=2,3,4,\cdots$). However, adding two extra transverse voltage probes  onto the same 1D ring lifts the minima of the longitudinal conductance to $G_L(Q_{\rm R}^{\rm min}) = I_4/(V_1-V_4) = e^2/h$. Nevertheless, the spin Hall conductance vanishes $G_{sH}^z(Q_{\rm R}^{\rm min}) \equiv 0$ at exactly these values 
of the SO coupling, while the amplitude of its quasiperiodic oscillations (which are not present in quantum spin-charge transport through simply-connected geometries~\cite{meso_spin_hall}) gradually decreases at 
large $Q_{\rm R}$ due to reflection at the ring-lead interface~\cite{purity}. 

\begin{figure}
\includegraphics[scale=0.75]{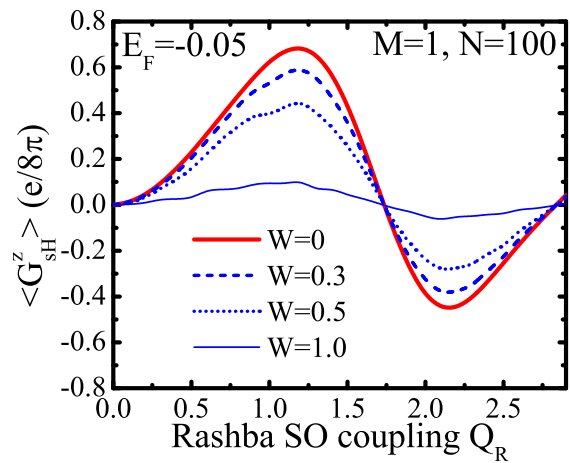}
\caption{The decay of the amplitude of disorder-averaged 
spin Hall conductance $\langle G^z_{\rm sH} \rangle$ with increasing strength 
$W$ of the disorder introduced in the same 1D ring whose ballistic transport 
regime is examined  in Fig.~\ref{fig:ring_1D}.}\label{fig:gsh_disorder}
\end{figure}

Finally, we examine the observability of spin Hall current in realistic rings of finite width and in the presence of spin-independent impurities~\cite{inoue}. Figure~\ref{fig:ring_2D} demonstrates that in 2D rings attached to four single channel probes the distinctive signatures---$G_{sH}^z(Q_{\rm R}^{\rm min})=0$ at specifically tuned (but harder to interpret~\cite{ac_ring}) $Q_{\rm R}^{\rm min}$---of quantum-interference dominated mesoscopic spin Hall effect can survive. When $M=2$, we observe that the frequency of $G_{sH}^z(Q_{\rm R})$ oscillations is almost doubled. This is due to the presence of the second harmonics in the ring, which is a well-known effect in the AB rings with large radius/width ratio~\cite{abring}. At larger widths, the quasi-periodicity of the $G_{\rm sH}$($Q_{\rm R})$ is destroyed since accumulated AC phases average over many Feynman paths through the ring, thereby "dephasing" visibility  of spin-interference effects~\cite{ac_ring}. When spin-independent scattering of static impurities occurs in disordered 1D rings, the amplitude of $G_{sH}^z(Q_{\rm R})$ in Fig.~\ref{fig:gsh_disorder} is reduced with increasing disorder strength $W$ of $V_{\rm dis} \neq 0$ in Eq.~(\ref{eq:rashba_hamil}),  simulated here by introducing a uniform random variable  $\varepsilon_{\bf m} \in [-W/2,W/2]$ at each lattice site ${\bf m}$.

{\em Conclusion}---We predict that pure spin Hall current dominated by {\em quantum-interference}  
effects will be generated in mesoscopic ring-shaped 1D and 2D conductors and, in principle, 
could be observed by measuring its {\em unequivocal} experimental signature---quasi-oscillatory 
pattern of the SO coupling dependent voltage~\cite{guo_spin_current,ewelina} induced by the spin 
flow exiting  from the Rashba spin-split  {\em multiply-connected} region through the single-open-channel 
electrodes.

\begin{acknowledgments}
We are grateful to E.I. Rashba and J. Sinova for immensely valuable insights.
This research was supported in part by ACS grant No. PRF-41331-G10.
\end{acknowledgments}



\end{document}